\documentclass[10pt]{article}

\usepackage{amsmath}
\usepackage{amssymb}

\usepackage{graphicx}

\usepackage{color}

\usepackage{float}
\usepackage[superscript,nomove]{cite}

\usepackage[labelfont=bf,labelsep=period,justification=raggedright]{caption}

\makeatletter
\renewcommand{\@biblabel}[1]{\quad#1.}
\makeatother

\date{}

\begin{document}

\begin{flushleft}
{\Large
\textbf{Agent-based model with multi-level herding for complex financial systems}
}
\\
Jun-Jie Chen$^{1,2}$,
Lei Tan$^{1,2}$,
Bo Zheng$^{1,2\ast}$
\\
$^{1}$Department of Physics, Zhejiang University, Hangzhou $310027$, China, $^{2}$Collaborative Innovation Center of Advanced Microstructures, Nanjing University, Nanjing 210093, China
\\
$^{\ast}$ E-mail: zhengbo@zju.edu.cn
\end{flushleft}

\section*{Abstract}
In complex financial systems, the sector structure and volatility clustering are respectively important features of the spatial and temporal correlations. However, the microscopic generation mechanism of the sector structure is not yet understood. Especially, how to produce these two features in one model remains challenging. We introduce a novel interaction mechanism, i.e., the multi-level herding, in constructing an agent-based model to investigate the sector structure combined with volatility clustering. According to the previous market performance, agents trade in groups, and their herding behavior comprises the herding at stock, sector and market levels. Further, we propose methods to determine the key model parameters
from historical market data, rather than from statistical fitting of the results. From the simulation, we obtain the sector structure and volatility clustering, as well as the eigenvalue distribution of the cross-correlation matrix, for the New York and Hong Kong stock exchanges. These properties are in agreement with the empirical ones. Our results quantitatively reveal that the multi-level herding is the microscopic generation mechanism of the sector structure, and provide new insight into the spatio-temporal interactions in financial systems at the microscopic level.

\newpage

Financial markets are complex systems with
many-body interactions. In recent years, large amounts
of historical financial data have sparked the interest of
scientists in various fields, including physicists, to quantitatively investigate the properties of the markets.
With physical concepts and methods, plenty of results have been obtained \cite{man95,gop99,liu99,bou01,tin01,kra02,gab03,sor03,bon03,joh03,qiu06,she09,zha11,pre11,fen12,pre13,che13,jia14,men14}.

From the view of physicists, the dynamic behavior and community structure of complex financial systems can be characterized by temporal and spatial correlation
functions. In stock markets, it is well-known that the volatilities are long-range
correlated in time, which is the so-called volatility clustering \cite{gop99,kra02,gab03,liu99,din93}. As to higher-order time correlations, it is discovered that the correlation between past returns and future volatilities is negative for almost all the stock markets in the world \cite{bla76,bou01,qiu06,glo93,eng93,zak94}, while currently the correlation is found to be positive only for the Chinese stock market \cite{qiu06,she09a}. In other words, the positive and negative returns influence the volatilities asymmetrically, which is known as the volatility asymmetry. During the financial crisis, the volatility asymmetry experiences local minima for developed economies, while reaches local maxima for transition economies \cite{ten10}. The spacial structure of the stock markets is explored by investigating the cross-correlation of stocks \cite{erb94,sol96,lal99,gop01,ple02,uts04,pan07,she09,pod10,jia12,jia14}.
With the random matrix theory (RMT), communities can be identified, which are usually associated with business sectors \cite{ple02,uts04,pan07}. The cross-correlation
matrix $C$ of price returns is analyzed to investigate the interactions between stocks \cite{lal99,gop01,ple02,uts04,pan07,she09,pod10,jia12,jia14}. The largest eigenvalue of
$C$ deviates significantly from the theoretical
distribution of the Wishart matrix, which is the cross-correlation matrix of non-correlated time
series. This eigenvalue represents the market mode, i.e., the price co-movement of the entire market, and the components of the corresponding eigenvector is relatively uniform for all stocks. For developed markets,
each eigenvector of other large eigenvalues is dominated by the stocks in a certain business sector \cite{gop01,ple02,jia14}. These large eigenvalues stand for the sector modes.

The spatio-temporal correlations are theoretically crucial in the understanding of the price dynamics, and practically useful
for the optimization of the investment portfolio. The sector structure and volatility clustering are, respectively, important features of the spatial and temporal correlations, which we focus on in this paper. In recent years, various models have been proposed to study volatility clustering with certain success \cite{lux00,gia01,cha01,kra02,ren06,fen12}, but the models for the sector structure are phenomenological and usually without interactions of investors \cite{ma04,she09}. On the other hand, although many activities have been devoted to the sector structure, its microscopic generation mechanism is not yet understood. Both the sector structure and volatility clustering are important characteristics of stock markets, and it remains challenging how to produce these two properties in one model.

As a powerful simulation technique, agent-based modeling is
widely applied in various fields \cite{gia01,cha01,bon02,eba04,ren06a,far09,sch09}. Recently, an agent-based model
is proposed for simulating the cumulative distribution of
returns and volatility clustering in stock markets \cite{fen12}. The concept in constructing the model
is to determine the key parameters from empirical data instead of
setting them artificially. In this paper, we construct an agent-based
model with a novel interaction mechanism, i.e., the multi-level herding, to investigate the sector structure combined with volatility clustering. Further,
we propose methods to determine the key model parameters from historical
market data rather than from statistical fitting of the results.

\section*{Results}

In stock markets, the
temporal evolution of stock prices and interactions between stocks are complicated. The price dynamics of a market naturally contains that of each individual stock. Recent research has reported that the price
dynamics of a market
can be decomposed into different modes of motion, such as the market mode and sector mode \cite{gop01,ple02,jia14}. The market mode is driven by interactions that are
common for all stocks in the market, and the sector mode is dominated by interactions of stocks within a sector. Therefore, the price dynamics of a market is multi-level. In financial markets, herding is one of the collective behaviors \cite{egu00,con00,hwa04,zhe04,ken11}. Investors cluster into groups when making
decisions, and these groups can be large. Since investors' herding behavior is essential to the price dynamics, it may be multi-level as well. In our model, we suppose
that agents' herding is composed of three different levels, that is, herding at stock, sector and market levels.

\textbf{Multi-level herding.} Our model is constructed based on the agents' daily trading,
i.e., buying, selling and holding stocks. In the model, there are
$N$ agents, $n$ stocks and $n_{sec}$ sectors. Each sector contains $n/n_{sec}$ stocks. Every agent holds only one stock, which is
randomly chosen from the $n$ stocks. In
a real market, an investor may hold different stocks. For simplicity, we suppose a reasonable
investor would trade his stocks separately, according to the performance
of each stock, even if his operation is based on an investment portfolio. Thus, the scenario for one investor holding, e.g., three stocks is basically the same as that for three investors with each holding one stock.

The stock price of the $i$-th stock on day $t$ is denoted by $Y_{i}(t)$,
and the logarithmic price return is $R_{i}(t)=\ln [Y_{i}(t)/Y_{i}(t-1)]$. Since investors' trading decisions in a real market is based on the previous stock performance
of different time scales, the investment horizon is introduced in our model for better description of agents' behavior. It has been reported in ref.~15
that the relative portion $\xi_{l}$ of investors with a $l$ days investment
horizon follows a power-law decay, $\xi_{l}\varpropto l^{- 1.12}$. The maximum investment horizon is denoted by $L$. With the condition of $\sum_{l=1}^{L}\xi_{l}=1$,
we normalize $\xi_l$ to be $\xi_{l}=l^{-1.12}/\sum_{l=1}^{L}l^{-1.12}$. Agents' trading decisions are made according to the past price returns. On day $t+1$, for an agent holding stock $i$ with a investment
horizon of $l$ days, $\sum_{m=0}^{l-1}R_{i}(t-m)$ represents the basis for estimating the previous stock performance. We introduce a weighted average return $R_{i}'(t)$
to describe the basis of all agents holding stock $i$. Since $\xi_{l}$ is the weight of $\sum_{m=0}^{l-1}R_{i}(t-m)$, $R_{i}'(t)$ is defined as
\begin{equation}
R_{i}'(t)=k\cdot\sum_{l=1}^{L}\left[\xi_{l}\sum_{m=0}^{l-1}R_{i}(t-m)\right].\label{eq:IH}
\end{equation}
We set the coefficient $k=1/(\sum_{l=1}^{L}\sum_{m=l}^{L}\xi_{m})$ to ensure that the fluctuation scale of $R_{i}'(t)$ is consistent with the one of $R_{i}(t)$, i.e.,  $|R_{i}'(t)|_{max}=|R_{i}(t)|_{max}$ (see Supplementary Information S1). If $L=1$, $R_{i}'(t)$ is just identical to $R_{i}(t)$.

\begin{figure}[H]
\begin{center}
\includegraphics[height=2.3in]{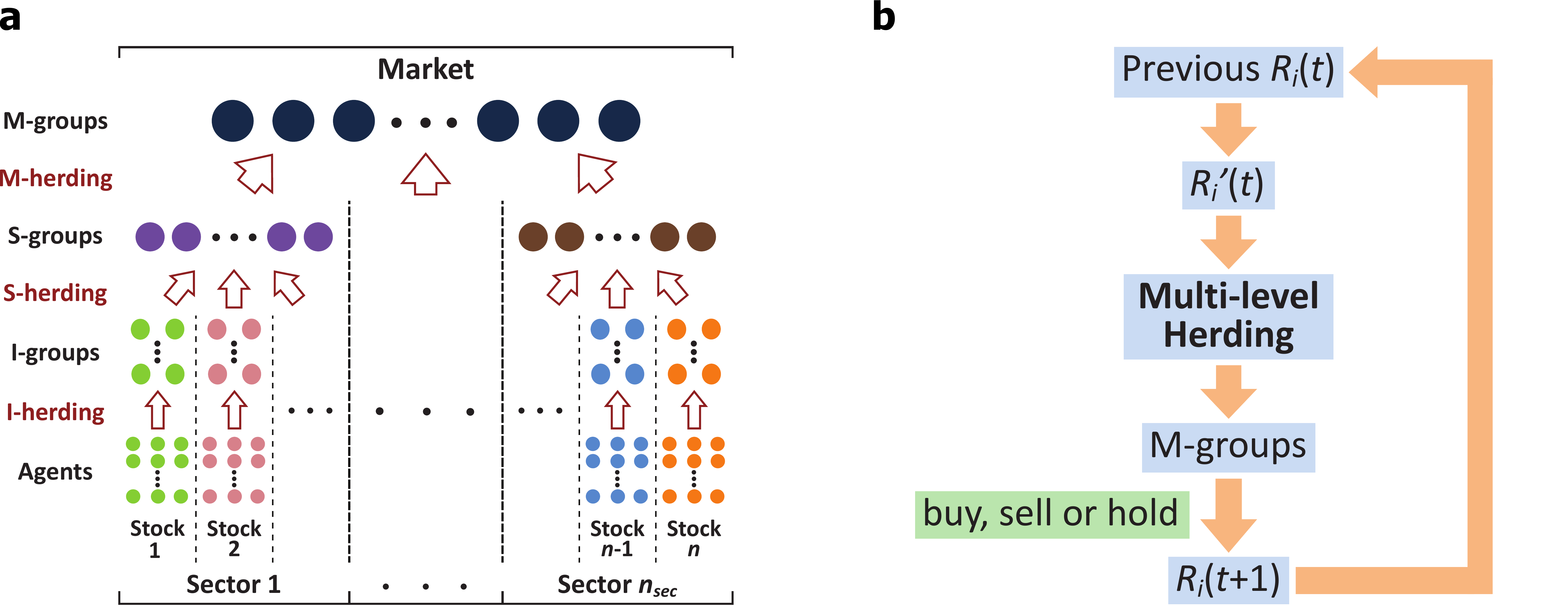}
\end{center}
\caption{\textbf{The schematic diagram of (a) the multi-level herding; (b) the procedure of simulation.} (a) ``I-herding'', ``S-herding'' and ``M-herding'' denotes the herding at stock level, sector level and market level, respectively.
}
\label{fig:Disp}
\end{figure}

In the model, agents' herding behavior comprises the herding at stock, sector and market levels. For convenience, we denote an agent holding stock $i$ by ``agent in stock $i$''. If stock $i$ belongs to sector $j$, this agent can be denoted by ``agent in sector $j$'' as well. A group formed by agents in stock $i$ or by agents in sector $j$ is respectively denoted by ``group in stock $i$'' or ``group in sector $j$''. The schematic diagram of the multi-level herding is displayed in Fig.~\ref{fig:Disp}(a). The agents in
a same stock first cluster into groups. This herding behavior at stock level is similar to the herding in other models which simulate only one stock. In a real market, the stocks in a same sector share the characteristics of the sector. Thus in our model, the groups in each sector further form larger groups, which is the herding at sector level. At last, the groups formed at sector level cluster into even larger ones, since all sectors share common features of the whole market. This is the herding at market level.

(i) Herding at stock level. The agents in
each individual stock first cluster into groups, which are called I-groups. We introduce
a herding degree $D^{I}$ to quantify the herding behavior
at this level. On day $t$, the herding degree
for the $i$-th stock is
\begin{equation}
D_{i}^{I}(t)=\bar{n}_{i}(t)/N_{i},
\end{equation}
where $\bar{n}_{i}(t)$ denotes the average number of agents in each I-group, and $N_{i}$ is the number of agents in the
$i$-th stock. Agents' herding behavior is based on their estimation of the previous stock performance. Since agents' basis for estimation on day $t$ is $|R_{i}'(t-1)|$, we set $\bar{n}_{i}(t)=|R_{i}'(t-1)|$. Thus,
\begin{equation}
D_{i}^{I}(t)=|R_{i}'(t-1)|/N_{i}.\label{eq:HaI}
\end{equation}
In the $i$-th stock, the number of I-groups
is $N_{i}/\bar{n}_{i}(t)=1/D_{i}^{I}(t)$, and
the agents randomly join in one of the
I-groups. After the herding at stock level for all the $n$ stocks,
the number of I-groups in the $j$-th sector and in the whole market
are, respectively, denoted by $N_{j}^{I}(t)$ and $N_{M}^{I}(t)$,
\begin{equation}
\left\{ \begin{array}{c}
N_{j}^{I}(t)=\sum\limits_{i\in j}[1/D_{i}^{I}(t)]\\
N_{M}^{I}(t)=\sum\limits_{i}[1/D_{i}^{I}(t)]
\end{array}.\right.
\end{equation}
Here $i\in j$ represents the stock $i$ in sector $j$.

(ii) Herding at sector level. The stocks in a same sector share the characteristics of the sector. At this level, agents' herding behavior is driven by the price co-movement of the sector, i.e., the prices of stocks in a sector tend to rise and fall at the same time. Thus the I-groups in a same sector would further form
larger groups, which are called S-groups. Here we introduce $H_{M}$ and $H_{j}$ to characterize the price co-movement degrees for stocks in the whole market and in sector $j$, respectively.
For the $j$-th sector, the average number of I-groups in each S-group is set
to be $n\cdot(H_{j}-H_{M})$, which represents the pure price co-movement of the
sector. Therefore the herding degree is
\begin{equation}
D_{j}^{S}(t)=n\cdot(H_{j}-H_{M})/N_{j}^{I}(t).\label{eq:HaS}
\end{equation}
In sector $j$, the number of S-groups is $1/D_{j}^{S}(t)$, and each I-group joins in one of the S-groups.

(iii) Herding at market level. Agents' herding behavior at this level is driven by the price co-movement of the entire market. The S-groups in different sectors share common features of the whole market, and thus cluster into larger groups. These groups are called M-groups. In the model, both the herding degrees at sector and market levels are computed based on the I-groups. The co-movement degree $H_{j}$ represents the percentage of connected I-groups, i.e., I-groups which co-move with each other. $\bar{H}=(\sum_{j}H_{j}/n_{sec})$ stands for the average percentage of connected I-groups. Since the group formation at this level is driven by the price co-movement of the whole market, we suppose that the number of connected I-groups should be the same for different sectors and equal to $\bar{H}\cdot N_{M}^{I}(t)$. We denote $\bar{H}\cdot N_{M}^{I}(t)/H_{j}$ by $N_{j}^{M}(t)$, and call it the effective number of I-groups for the $j$-th sector. $N_{j}^{M}(t)$ satisfies $H_{j}\cdot N_{j}^{M}(t)=\bar{H}\cdot N_{M}^{I}(t)$. On the other hand,
$n\cdot H_{M}$ represents the price co-movement for stocks in the whole market. Thus for the S-groups in sector $j$, the herding degree at market level is
\begin{equation}
D_{j}^{M}(t)=n \cdot H_{M}/N_{j}^{M}(t),
\end{equation}
and the number of M-groups is $1/D_{j}^{M}(t)$. The total number of M-groups in the market is the maximum of $1/D_{j}^{M}(t)$ for different $j$. With all M-groups numbered, an S-group in sector $j$ joins in one of the first $1/D_{j}^{M}(t)$ M-groups.

In the formation of S-groups, the I-groups in a
same stock tend not to join in a same S-group, otherwise these I-groups would have gathered together during the herding
at stock level. Similarly, in the formation of M-groups, the S-groups in a
same sector tend not to join in a same M-group.
In other words, an I-group prefers to join in an
S-group with no other I-groups from the same stock, and an S-group
prefers to join in an M-group with no other S-groups from the same
sector.

After the herding for the three levels, all agents cluster
into M-groups. Since intraday trading is not persistent in empirical trading
data \cite{eis07}, we suppose that each day only one trading decision is made by every agent. The agents in a same M-group make
a same trading decision with a same probability. Considering each agent operates one share, we denote the decision
of the $\alpha$-th agent on day $t$ by
\begin{equation}
\phi_{\alpha}(t)=\begin{cases}
\:1 & \textnormal{buy}\\
-1 & \textnormal{sell}\\
\:0 & \textnormal{hold}
\end{cases},
\end{equation}
and the probabilities of buying, selling and holding decisions of M-groups
are denoted by $P_{buy}$, $P_{sell}$ and $P_{hold}$, respectively.
The same as the previous models \cite{fen12,che13}, we suppose that the buying and selling probabilities are equal, i.e., $P_{buy}=P_{sell}=P$, thus $P_{hold}=1-2P$.
The return of the $i$-th stock is defined as the difference
of the demand and supply of this stock, i.e., the difference between
the number of agents buying and selling the stock,
\begin{equation}
R_{i}(t)=\sum_{\alpha \in i}\phi_{\alpha}(t).\label{eq:R}
\end{equation}
Here $\alpha \in i$ represents the agent $\alpha$ in stock $i$.

\textbf{Determination of parameters $H_{M}$ and $H_{j}$.} The New York Stock Exchange (NYSE) and Hong Kong Stock Exchange (HKSE) are the two representative stock markets considered in this paper. The NYSE is one of the world's most mature markets in the West, and the HKSE is an important market in Asia. We collect the daily closing
price data of $150$ stocks in the NYSE and
HKSE, respectively (Methods).
For the comparison of different time series of returns, the normalized
return $r_{i}(t)$ is introduced,
\begin{equation}
r_{i}(t)=[R_{i}(t)-\langle R_{i}(t)\rangle]/\sigma,\label{eq:norm}
\end{equation}
where $\langle \cdots\rangle $ represents the average
over time $t$, and $\sigma=\sqrt{\langle R_{i}^{2}(t)\rangle -\langle R_{i}(t)\rangle ^{2}}$ is the standard deviation of
$R_{i}(t)$.

The parameters $H_{M}$ and $H_{j}$ are introduced to characterize the price co-movement degrees for stocks in the whole market
and in sector $j$, respectively. Actually, the price co-movement of stocks can be characterized by the similarities in the signs and amplitudes of the returns for different stocks. We denote
the number of stocks in a sector by $n_{s}$, thus $n_{s}=n/n_{sec}$. In the calculation of $H_{M}$, we simply set $n_{s}=n$. On each day $t$, according to the sign of $r_{i}(t)$, these
stocks are grouped into two market trends, i.e., the rising and the falling.
The amplitudes of the rising and falling trends on day $t$ are defined as
$v^{+}(t)$ and $v^{-}(t)$, respectively,
\begin{equation}
\left\{ \begin{array}{c}
v^{+}(t)=\sum_{i,r_{i}(t)>0}r_{i}^{2}(t)/n_{s}\\
v^{-}(t)=\sum_{i,r_{i}(t)<0}r_{i}^{2}(t)/n_{s}
\end{array}\right..
\end{equation}
The two trends are usually not in balance, and we suppose that these $n_{s}$
stocks are dominated by either of the two trends, according to the magnitudes of $v^{+}(t)$ and $v^{-}(t)$. The amplitude $v^{d}(t)$ of the dominating trend and the amplitude $v^{n}(t)$ of the non-dominating one
are
\begin{equation}
\left\{ \begin{array}{c}
v^{d}(t)=max\{v^{+}(t),\: v^{-}(t)\}\\
v^{n}(t)=min\{v^{+}(t),\: v^{-}(t)\}
\end{array}\right..
\end{equation}
We call the stocks grouped into the dominating trend the ``dominating
stocks''. The price co-movement is a common property for all the $n_{s}$
stocks, hence the total amplitude is $v^{d}(t)-v^{n}(t)$. Besides, we take into consideration the similarity in the signs of the returns for these stocks. This similarity is defined as the percentage of the dominating stocks. The number of the dominating stocks is denoted by $n^{d}(t)$,
and the percentage is $\zeta(t)=n^{d}(t)/n_{s}$. We take the average over time $t$ for $\zeta(t)$ and $v^{d}(t)-v^{n}(t)$, denoted respectively by $\left\langle \zeta(t)\right\rangle$ and $\left\langle v^{d}(t)-v^{n}(t)\right\rangle$.
Then, the co-movement degree $H_{M}$ and $H_{j}$ are
\begin{equation}
\left\{ \begin{array}{c}
H_{M}=\left.\left\langle \zeta(t)\right\rangle \cdot\left\langle v^{d}(t)-v^{n}(t)\right\rangle \right|_{\textnormal{market}}\\
H_{j}=\left.\left\langle \zeta(t)\right\rangle \cdot\left\langle v^{d}(t)-v^{n}(t)\right\rangle \right|_{\textnormal{$j$-th sector}}
\end{array}.\right.
\end{equation}
Here $|_{\textnormal{market}}$ and $|_{\textnormal{$j$-th sector}}$ represent the calculations for the stocks in the whole market and in the $j$-th sector, respectively. These parameters for the NYSE and HKSE are shown in Table~\ref{tab:value}.

\textbf{Simulation of the model.} The number of stocks is $n=150$ and the number of sectors is $n_{sec}=5$, which are the same as those of the empirical data we collect for the NYSE and HKSE (Methods). We set the number of agents $N$ to be $600,000$, and the maximum investment horizon $L$ to be $1000$ trading days (Methods). In our model, the investment horizons of $94$ percent of agents are shorter than $500$ days (Methods), which is similar to the previous research \cite{fen12}.  Estimated from the historical market data and investment report, the buying or selling probability is $P=0.363$ for the NYSE and $P=0.317$ for the HKSE (Methods). With $H_{M}$ and $H_{j}$ determined for the NYSE and HKSE respectively,
our model produces the time series $R_{i}(t)$ of each stock. The schematic diagram of the simulation procedure is displayed in Fig.~\ref{fig:Disp}(b):

Initially, the returns of the first $L$ time steps are set to
be $0$ for all the $n$ stocks. On day $t$,
we calculate agents' basis $R_{i}'(t)$ according to equation~(\ref{eq:IH}), and then $D_{i}^{I}(t)$ according to equation~(\ref{eq:HaI}) for each stock $i$.
The agents in the $i$-th stock randomly join in one of the $1/D_{i}^{I}(t)$
I-groups. Next,
the I-groups in the $j$-th sector join in one of the $1/D_{j}^{S}(t)$
S-groups, and then each S-group in the $j$-th sector joins in one of the
$1/D_{j}^{M}(t)$ M-groups. After the herding for these three levels, the agents in a same
M-group make a same trading decision (buy, sell or hold) with the
same probability ($P_{buy}$, $P_{sell}$ or $P_{hold}$). Thus the
returns for each stock on day $t$ are calculated with equation~(\ref{eq:R}). The groups disband after their decisions are made. Repeating
this procedure, we obtain the time series of returns for all the stocks in
the market. For equilibration, the first $10,000$ data points of returns are abandoned for each stock, and the length of the time
series from simulation is the same as that of the empirical data.

\textbf{Simulation results. }

\begin{figure}[H]
\begin{center}
\includegraphics[height=2.3in]{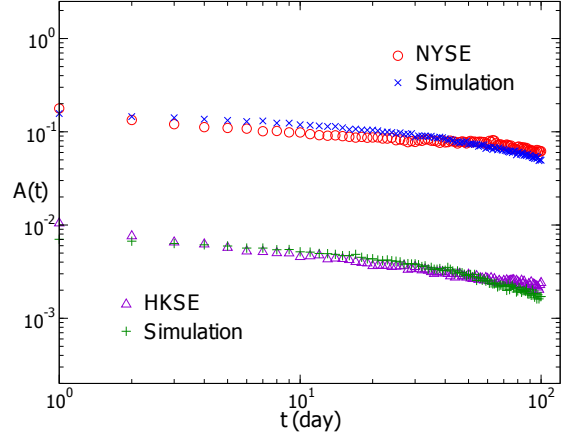}
\end{center}
\caption{\textbf{The average auto-correlation functions of volatilities for the NYSE and HKSE, and for the corresponding simulations.} For clarity, the curves for the HKSE have been shifted down by a factor of $20$.
}
\label{fig:A}
\end{figure}

From the calculation for the simulated returns, we obtain the sector structure and volatility clustering. For each stock $i$, the volatility clustering is characterized by the auto-correlation function of volatilities \cite{gop99,kra02}, which is defined as
\begin{equation}
A_{i}(t)=[\langle |r_{i}(t')||r_{i}(t'+t)|\rangle -\langle |r_{i}(t')|\rangle ^{2}]/A_{i}^{0}.
\end{equation}
Here $A_{i}^{0}=\langle |r_{i}(t')|^{2}\rangle -\langle |r_{i}(t')|\rangle ^{2}$, and $\langle \cdots\rangle $ represents the average
over time $t'$. Thus, the auto-correlation function of volatilities averaged over all stocks is $A(t)= \sum_{i} A_{i}(t)/n$. As shown in Fig.~\ref{fig:A}, for both the NYSE and HKSE, $A(t)$ for the simulations is in agreement with that for the empirical data.

\begin{figure}[H]
\begin{center}
\includegraphics[height=2.3in]{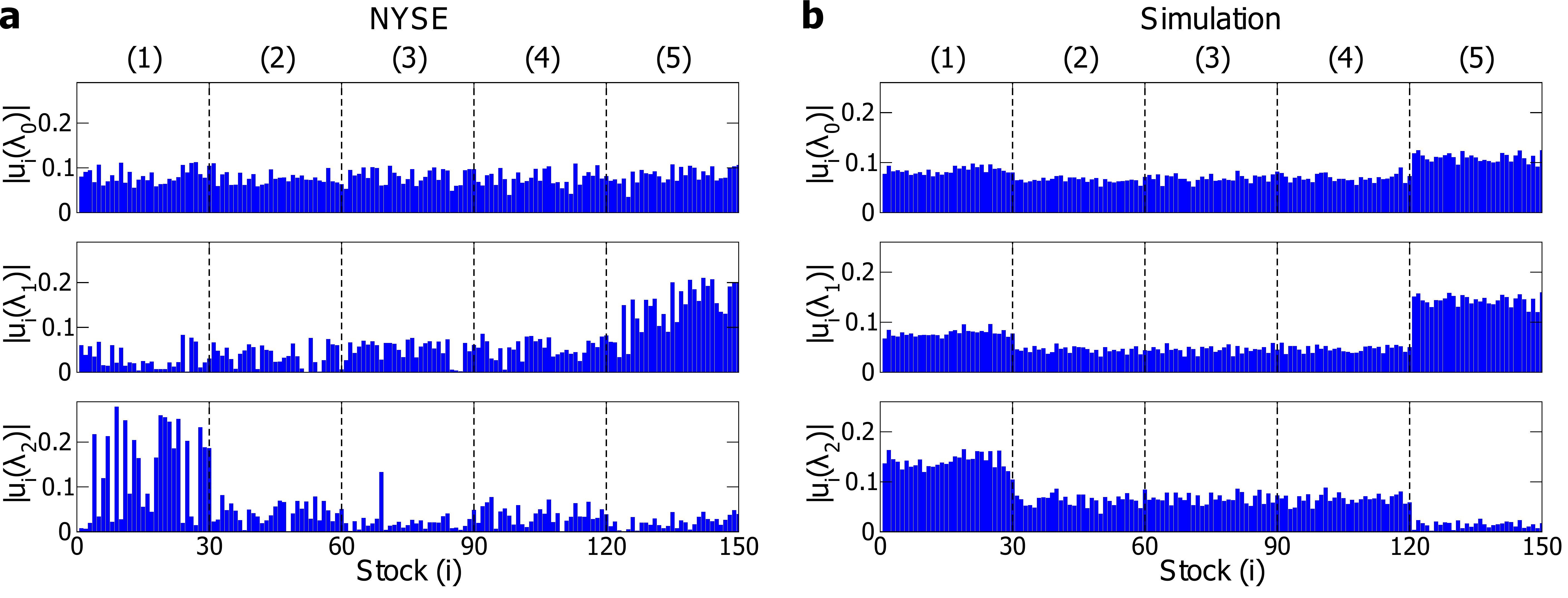}
\end{center}
\caption{\textbf{The absolute values of the eigenvector components $u_{i}(\lambda)$ corresponding to the three largest eigenvalues for the cross-correlation matrix $C$ calculated from (a) the empirical data in the NYSE; (b) the simulated returns for the NYSE.} Stocks are arranged according to business sectors separated by dashed lines. (1): Basic Materials; (2): Consumer Goods; (3): Industrial Goods; (4): Services; (5): Utility.
}
\label{fig:NYV}
\end{figure}

\begin{figure}[H]
\begin{center}
\includegraphics[height=2.3in]{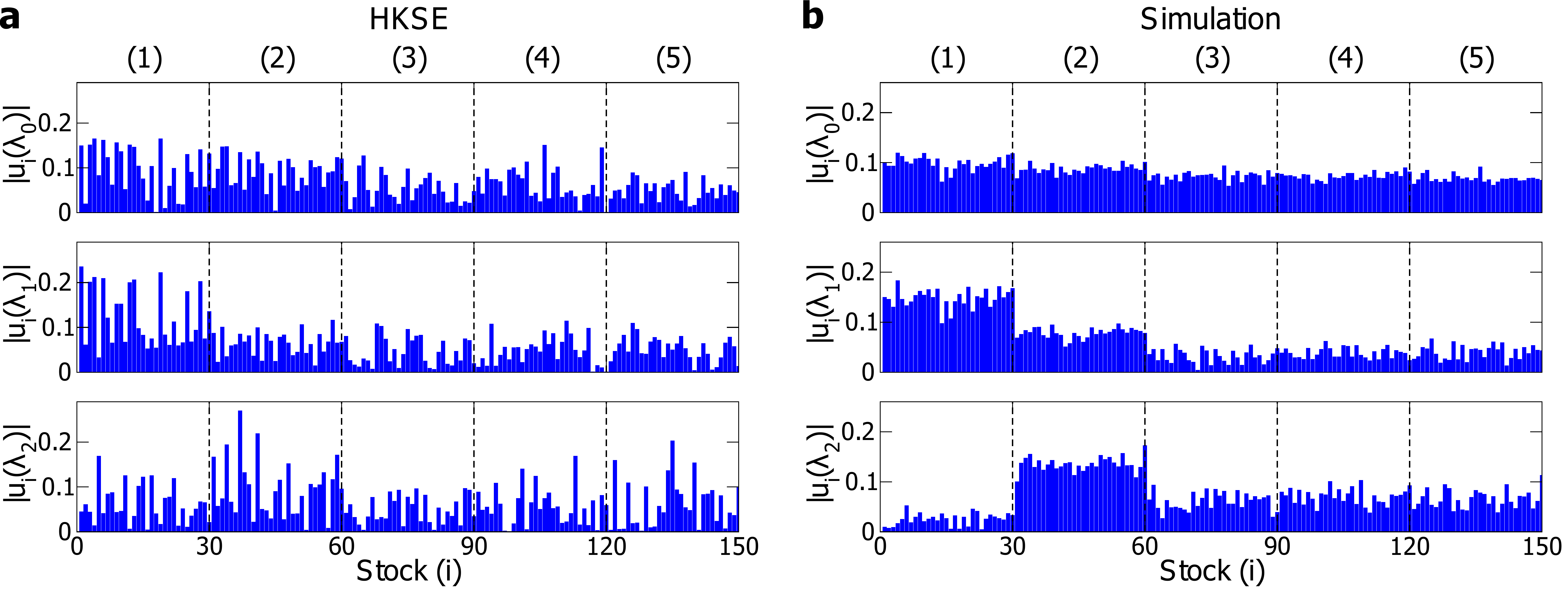}
\end{center}
\caption{\textbf{The absolute values of the eigenvector components $u_{i}(\lambda)$ corresponding to the three largest eigenvalues for the cross-correlation matrix $C$ calculated from (a) the empirical data in the HKSE; (b) the simulated returns for the HKSE.} Stocks are arranged according to business sectors separated by dashed lines. Sector $(2)$ and $(3)$ are composed
of two business sectors, respectively (Methods). (1): Real Estate Development; (2): Conglomerates - Industrial Goods; (3): Basic Materials - Technology; (4): Services; (5): Consumer Goods.
}
\label{fig:HKV}
\end{figure}

To characterize the spacial structure, we first compute the equal-time cross-correlation matrix $C$ \cite{lal99,she09,ple99}, of which each element is
\begin{equation}
C_{ij}=\langle r_{i}(t) r_{j}(t) \rangle.
\end{equation}
Here $\langle \cdots\rangle $ represents the average
over time $t$, and $C_{ij}$ measures the correlation between the returns of the $i$-th and $j$-th stocks. From the definition, $C$ is a real symmetric matrix with $C_{ii}=1$, and the values of other elements $C_{ij}$ are in the interval $[-1,1]$. The first, second and third largest eigenvalues of $C$ are denoted by $\lambda_{0}$, $\lambda_{1}$ and $\lambda_{2}$, respectively. Now we focus on the components $u_{i}(\lambda)$ of the eigenvector for the three largest eigenvalues. The empirical result of the NYSE is displayed in Fig.~\ref{fig:NYV}(a). For $\lambda_{0}$, the components of the corresponding eigenvector are relatively uniform. The eigenvectors of $\lambda_{1}$ and $\lambda_{2}$ are dominated by sector $(5)$ and sector $(1)$ respectively, with the components significantly larger than those in other sectors. These features are reproduced in our simulation, and the results are shown in Fig.~\ref{fig:NYV}(b). The empirical result of the HKSE is displayed in Fig.~\ref{fig:HKV}(a). The eigenvectors of $\lambda_{1}$ and $\lambda_{2}$ are respectively dominated by sector $(1)$ and sector $(2)$, and these features are simulated by our model, shown in Fig.~\ref{fig:HKV}(b). For the HKSE, the scenario is somewhat complicated \cite{ouy14}, since a company in the HKSE usually runs various business. As a result, the components of the eigenvector of $\lambda_{0}$ are not so uniform as those in the NYSE. The dominating sectors for $\lambda_{1}$ and $\lambda_{2}$ are less prominent, especially for $\lambda_{2}$.

\begin{figure}[H]
\begin{center}
\includegraphics[height=2.3in]{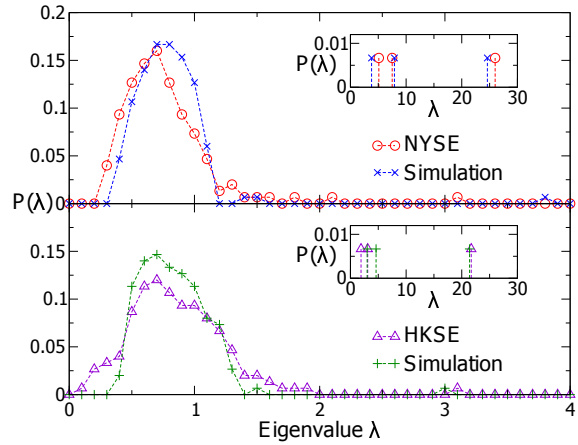}
\end{center}
\caption{\textbf{The probability distribution of the eigenvalues of the cross-correlation matrix $C$ for the NYSE and HKSE, and for the corresponding simulations.} The probability distribution of the three largest eigenvalues is shown in the inset. For the NYSE, the three largest eigenvalues are ($\lambda_{0}$, $\lambda_{1}$, $\lambda_{2}$) = (26.01, 7.45, 5.13), in comparison with (24.62, 7.93, 3.82) for the simulation. For the HKSE, the three largest eigenvalues are ($\lambda_{0}$, $\lambda_{1}$, $\lambda_{2}$) = (21.70, 3.06, 1.89), in comparison with (21.43, 4.57, 2.97) for the simulation.
}
\label{fig:lamda}
\end{figure}

Also, the distribution of the eigenvalues is calculated from the simulated returns. As displayed in Fig.~\ref{fig:lamda}, for the NYSE and HKSE, the bulk of the distribution of eigenvalues and the three largest eigenvalues from the simulation are in agreement with those from the empirical data.

\section*{Discussion}

In financial markets, the sector structure and volatility clustering are respectively important features of the spatial and temporal correlations. However, the microscopic generation mechanism of the sector structure is not yet understood. Especially, how to produce these two features in one model remains challenging.

To investigate the sector structure combined with volatility clustering, we construct an agent-based model with a novel interaction mechanism, that is, the multi-level herding. The model is based on the individual and collective behaviors of investors in real markets. According to the previous market performance, agents trade in groups, and their herding behavior comprises the herding at stock, sector and market levels. The key parameters, $H_{M}$ and $H_{j}$, are introduced to characterize the price co-movement degrees for stocks in the whole market
and in sector $j$, respectively. We propose methods to determine these parameters
from historical market data rather than from statistical fitting of the results. Other parameters $L$ and $P$ are also estimated from the empirical findings.

With parameters determined for the NYSE and HKSE respectively, our model produces the corresponding time series of returns. From these time series, we obtain the sector structure and volatility clustering, as well as the eigenvalue distribution of the cross-correlation matrix $C$. These properties are in agreement with the empirical ones. Our results quantitatively reveal that the multi-level herding is the microscopic generation mechanism of the sector structure, and provide new insight into the spatio-temporal interactions in financial systems at the microscopic level.
The mechanism of the multi-level herding, including the concept of characterizing the price co-movement with parameters $H_{M}$ and $H_{j}$, can also be applied to other complex systems with similar community structures.

\section*{Methods}

\textbf{Data.} Our data are obtained from ``Yahoo$!$ Finance'' (http://finance.yahoo.com). We collect the daily data of closing prices of $150$
large-cap stocks from $5$ business sectors in the NYSE, i.e., the Basic Materials, Consumer Goods, Industrial Goods, Services and Utility, with $30$ stocks
from each sector. The data are from Jan., $1990$ to Dec., $2006$ with $4286$ data
points for every stock. For comparison, we also collect the daily
closing price data of $150$ stocks in the HKSE to form $5$ sectors, with $30$ stocks in each sector, and most of these stocks are large-cap stocks. The data are from Jan, $2003$ to Sep., $2011$ with $2146$ data points for each stock. The sector structure of the HKSE
is somewhat complicated \cite{ouy14}, since a company in the HKSE usually runs various business. According to the dominating stocks of eigenvectors of the cross-correlation matrix $C$, the sectors in the HKSE are not so strict as those in the NYSE, and may be composed of two business sectors. Specifically, the second sector comprises $14$ stocks from the Conglomerates and $16$ stocks from the Industrial Goods. The third sector consists of $12$ stocks from the Basic Materials and $18$ stocks from the Technology. The stocks in other three sectors are, respectively, from the Real Estate Development, Services and Consumer Goods.

\textbf{Parameter $N$ and $L$.} To simulate the properties of a real stock market, the number of agents $N$ should not be too small, since we suppose that every agent operates only one share of a stock. We set $N=600,000$, i.e., $4000$ agents holding a same stock on average. The simulation results are not sensitive to $N$. For example, we obtain almost the same results for $N=450,000$ or $N=750,000$. $N$ has no influence on the sector structure and the eigenvalue distribution of the matrix $C$, and slightly affects the amplitude of the auto-correlation function $A(t)$ of volatilities averaged over stocks. The investment horizons of investors range from one day to more than one year \cite{men10}. Considering there could be some investors with long investment horizon in the stock market, we set $L=1000$ trading days in our model. Thus according to the relative portion of investors, the investment horizons of $94$ percent of agents are shorter than $500$ days, suggesting that most agents do not estimate previous market performance over a too long time period. $L$ affects the temporal and spatial properties little. Similar with $N$, the simulation results are almost the same for, e.g., $L=800$ or $L=1200$.

\textbf{Determination of parameter $P$.} We first determine the daily buying, selling and holding
probabilities of a single investor in a real market, which is denoted by $p_{buy}$, $p_{sell}$ and $p_{hold}$ respectively. $p_{buy}$
and $p_{sell}$ are supposed to be equal, i.e., $p_{buy}=p_{sell}=p$.
The time series of returns of the $150$
stocks in the NYSE are from Jan., $1990$ to Dec., $2006$. According
to ``The $2010$ Institutional Investment Report'' \cite{ton10}, the average percentage
of institutional holdings of shares in the top 1000 U.S. corporations during these years
is $60.3$ percent (see Supplementary Information S2). There are two kinds of investors in stock markets, that is individual and institutional investors. The percentage of holding shares
for the individual investors is thus $39.7$ percent.
According to ref.~15, the yearly average ratio between the number of shares an investor trades and the number of shares he holds is $1.64$. This ratio corresponds to the yearly average trading times of an investor.
Since institutional investors contribute little of the trading times \cite{fen12}, we ignore
their trades. So the
yearly trading times for an individual investor is $1.64/0.397=4.13$.
Since there are $250$ trading days in every year, the daily trading probability
is $4.13/250=p_{buy}(t)+p_{sell}(t)=2p$. Therefore,
$p$ is $0.00826$ for the NYSE. For the HKSE, the corresponding empirical data are not available to us, and we assume that $p$ is $0.00826$ as well.

The agents in a same M-group are connected. We suppose that if one agent
in the group decides to buy or sell the stock, the whole group would make the same decision. In the model, the average number of agents in an M-group is $n\cdot H_{M}$.
Therefore, the buying or selling probability of an M-group is $P=1-(1-p)^{n\cdot H_{M}}$. Thus, $P$ is $0.363$ for the NYSE and $0.317$ for the HKSE.


\section*{Acknowledgments}
This work was supported in part by NNSF of China under Grant Nos. 11375149 and
11075137, and Zhejiang Provincial Natural Science Foundation of China under Grant No.
Z6090130.

\section*{Author contributions}
J.J.C., L.T. and B.Z. conceived the study; J.J.C. and L.T. designed and performed the
research; J.J.C. and L.T. performed the statistical analysis of the data; J.J.C., L.T. and B.Z. discussed the results and contributed to the
text of the manuscript.

\section*{Additional information}
\textbf{Competing financial interests:} The authors declare no competing financial interests.

\begin{table}[H]
\caption{\textbf{The values of parameters $H_{M}$ and $H_{j}$ for the NYSE and HKSE.} $H_{M}$ and $H_{j}$ are introduced to characterize the price co-movement degrees for stocks in the whole market
and in sector $j$, respectively. We determine these parameters from the historical market data for each stock exchange.
}
\begin{tabular}{c|c|ccccc}
\hline
 & $H_{M}$ & $H_{1}$ & $H_{2}$ & $H_{3}$ & $H_{4}$ & $H_{5}$\tabularnewline
\hline
NYSE & 0.363 & 0.491 & 0.414 & 0.438 & 0.431 & 0.546\tabularnewline
HKSE & 0.306 & 0.426 & 0.406 & 0.364 & 0.361 & 0.340\tabularnewline
\hline
\end{tabular}
\label{tab:value}
\end{table}

\end{document}